\documentclass[twocolumn,twocolappendix]{openjournal}

\newcommand{\lya}{Ly$\alpha$~}

\usepackage{xcolor}
\usepackage{textgreek}
\usepackage{amsmath}
\usepackage[utf8]{inputenc}
\usepackage[english]{babel}

\usepackage{hyperref}
\hypersetup{
    unicode, 
    colorlinks=true,
    linkcolor=linkcolor,
    citecolor=linkcolor,
    filecolor=linkcolor,
    urlcolor=linkcolor,
}
\usepackage{color,colortbl}
\definecolor{linkcolor}{rgb}{0.0,0.3,0.5}
\usepackage{tensind}
\tensordelimiter{?}
\DeclareGraphicsExtensions{.bmp,.png,.jpg,.pdf}
\usepackage{verbatim}
\usepackage[normalem]{ulem}
\usepackage{orcidlink}
\usepackage{soul}
\begin{document}
\title{A multi-eigenbasis approach to covariance matrix denoising for cosmological inference}

\author{Wynne Turner\orcidlink{0009-0008-3418-5599}}
\email{turner.1839@osu.edu}
\affiliation{Department of Astronomy, The Ohio State University, 4055 McPherson Laboratory, 140 W 18th Avenue, Columbus, OH 43210, USA}
\affiliation{Center for Cosmology and AstroParticle Physics, The Ohio State University, 191 West Woodruff Avenue, Columbus, OH 43210, USA}

\begin{abstract}
    Accurate covariance matrix estimation is crucial to cosmological analyses, enabling unbiased parameter inference with well-calibrated uncertainties. Obtaining a reliable estimate generally requires far more independent samples than the dimension of the data vector, which is not always feasible. This challenge is especially relevant for the 3D \lya forest analysis, which measures the \lya auto-correlation and its cross-correlation with quasars in bins of comoving separation to jointly constrain cosmological parameters. The consequence is a very large data vector, and the data-driven covariance measured from sub-samples is non-invertible. The current approach applies a smoothing procedure to the off-diagonals of the correlation matrix to establish invertibility, but this does not fully capture the true correlation structure. In this work, I present a novel multi-eigenbasis denoising method for the data-driven covariance matrix, developed in the context of the 3D \lya forest analysis and conditioned on DESI DR1 mock simulations. The measured noisy covariance is first projected onto the eigenbasis of a mock-based reference, yielding an initial denoised estimate. A weighted residual correction is then constructed by projecting the noisy residual onto a second eigenbasis derived from a mock-trained classifier, capturing correlation structure not recovered in the initial reconstruction. I validate the method on mock covariances withheld from classifier training and find significant improvements over the current smoothing-based approach in matrix-level reconstruction metrics and in the recovery of cosmological parameter posteriors when compared to those obtained from the true covariance measured from many mock realizations.
\end{abstract}


\maketitle

\section{Introduction}
\label{sec:intro}
The covariance matrix quantifies the statistical uncertainties of the data points in an analysis and their correlations. It is therefore of critical importance to large-scale structure surveys, which often aim to constrain cosmological parameters to sub-percent precision. As surveys seek to improve their constraining power by extending model fits into the nonlinear regime and measuring additional summary statistics such as cross-correlations with new tracers, the growing dimensionality of the data vector makes accurate covariance estimation an increasingly pressing challenge. When an analytic solution is not feasible, the sample covariance can be estimated through several approaches.

A common method is to divide the dataset into sub-samples and apply resampling techniques such as bootstrap or jackknife estimation. A central concern with this approach is that the assumption of sub-sample independence may be violated due to large-scale correlations present in the data. Additionally, it can be difficult or even impossible to obtain a sufficiently large number of sub-samples relative to the size of the data vector. Other approaches rely on simulated datasets, alleviating this concern provided a sufficiently large number of mock realizations is available. However, a covariance matrix estimated purely from mocks requires that those mocks capture all features present in the true data covariance and do not introduce a systematic bias. Further, the limited number of samples $N$ used to estimate the covariance relative to the data vector dimension $n$ introduces noise that biases the inverse covariance \citep{hartlap_2007} and increases inferred uncertainties \citep{dodelson_schneider_2013,taylor_joachimi_kitching_2013}. Several works have further studied the propagation of this noise into cosmological parameter estimation \citep[e.g.,][]{percival_covariance_2014,taylor_joachimi_2014,sellentin_heavens_2016,percival_2022}. However, these corrections are approximations that diminish in accuracy as $N$ approaches $n$. When $N < n$, the Wishart-distributed sample covariance matrix becomes singular and therefore cannot be inverted for use in cosmological inference.

Several methods have been proposed in the literature to mitigate these concerns by ``denoising'' the noisy sample covariance matrix. The linear shrinkage estimator \citep{pope_szapudi_linear_shrinkage_2008} tunes the sample covariance toward a target matrix. Nonlinear shrinkage estimators \citep[e.g.,][]{Ledoit_Wolf_2012,lam2016nonparametric,joachimi_2017} adopt a more sophisticated approach, applying individual, data-driven transformations to the eigenvalues of the sample covariance while retaining the sample eigenvectors. See \cite{looijmans_2025_shrinkage_comparison} for a recent review of shrinkage estimators in a cosmological context. Similarly, \cite{farina_2026_RIE} recently proposed the application of rotational invariant estimators to denoise the large-scale structure covariance matrix. While these methods are very promising, in the case of a singular sample covariance matrix, $(N-n)$ eigenvalues are identically zero and their associated eigenvectors are entirely uninformative. There have also been recent developments in using neural networks to denoise the sample covariance matrix \citep[e.g.,][]{desanti_2022}, though training such models may be computationally prohibitive for very large $n$.

Obtaining a reliable estimate of the \lya forest covariance matrix is especially challenging. The \lya forest traces the distribution of neutral hydrogen in the intergalactic medium through absorption lines in quasar spectra. The 3D \lya analysis measures the auto-correlation of this absorption and its cross-correlation with quasar positions in bins of comoving separation along ($r_\parallel$) and across ($r_\perp$) the line of sight \citep[e.g.,][]{slosar_2011,busca_2013_lya_boss,kirkby_2013,slosar_2013_lya_boss,fontribera_2013,fontribera_2014_lyaqso_boss,delubac_boss_dr11,bautista_2017,dmdb_2017,de_sainte_agathe_baryon_2019,blomqvist_baryon_2019,bourboux_completed_2020,desi_y1_4,desi_dr2_lya}. These correlations were first measured with the Baryon Oscillation Spectroscopic Survey \citep[BOSS;][]{dawson_boss_2013}, with covariance estimation techniques ranging from an analytic Wick expansion \citep[see e.g.,][]{slosar_2011,fontribera_2014_lyaqso_boss,delubac_boss_dr11} to data-driven approaches using sub-samples of comparable statistical size \citep[e.g.,][]{busca_2013_lya_boss,slosar_2013_lya_boss,delubac_boss_dr11,bautista_2017,dmdb_2017}. \cite{delubac_boss_dr11} was the first to compute the covariance over the full $(r_\parallel,r_\perp)$ coordinate grid, making the matrix non-invertible with $N<n$. They therefore introduced a smoothing procedure that averages all off-diagonal elements of the correlation matrix sharing the same separations along and across the line of sight, effectively reducing it to a function of $(\Delta r_\parallel,\Delta r_\perp)$ only. This smoothing approach has been employed in all subsequent \lya 3D analyses.

Later, correlation function measurements with the extended-BOSS \citep[eBOSS;][]{dawson_eboss_2016} used sub-samples defined by \texttt{HEALPix} pixels \citep{healpix_2005} to estimate the covariance and apply the smoothing procedure. More recently, these measurements have been performed with the Dark Energy Spectroscopic Instrument \citep[DESI;][]{desi_collaboration_desi_2016,desi_instrument_overview_2022}, including baryon acoustic oscillation (BAO) measurements from Data Release 1 \citep[DR1;][]{desi_y1_4} and DR2 \citep{desi_dr2_lya}, as well as a full-shape measurement from DR1 \citep{cuceu_2025}. These analyses all use the same \texttt{HEALPix}-based sub-sampling and smoothing of the correlation matrix described above, but were the first to account for the cross-covariance between the \lya auto- and cross-correlations, finding that assuming independence leads to a $\sim10\%$ underestimation of the BAO uncertainties \citep{desi_y1_4}. While the long-standing smoothing procedure has been validated, most recently in the context of BAO \citep{cuceu_desi_lya_validation} and the Alcock-Paczy\'{n}ski (AP) parameter from the full-shape correlation function \citep{cuceu_2025}, the process of averaging off-diagonal elements smooths out large-scale correlation structure present in the measured covariance. One potential alternative to bypass this limitation is to compress the data vector to obtain a well-conditioned sample covariance, as explored in the context of the \lya forest by \cite{gerardi_2024_compression}, though such methods typically still require a mock-based covariance to determine the optimal compression framework. Compressing the data vector by performing the analysis in multipoles may offer another promising direction, though recent results suggest this compression leads to a degradation in constraining power on the BAO parameters \citep{karacayli_2026_multipoles}.

This work introduces a new framework for denoising the data-driven sample covariance matrix. Given a well-conditioned mock-based reference correlation matrix, I project the noisy correlation matrix onto the eigenbasis defined by the reference matrix. The resulting diagonal elements, or pseudo-eigenvalues, are used to construct an initial denoised estimate. A residual correction then adds small corrections to this reconstruction to account for the fact that the reference correlation matrix cannot be fully representative of the truth. I construct a second eigenbasis from multiple residual matrices computed from different mock suites and train a weighted distance classifier to classify the noisy correlation matrix. The resulting weights are used to down-weight mock suites with less similarity to the data and project the noisy residual onto the appropriately weighted residual eigenbasis. This residual reconstruction is added to the initial denoised matrix to produce a final denoised correlation matrix. I then use the measured variance of the input covariance to rescale the denoised correlation matrix back to its covariance, preserving the statistical interpretation of the data.

This paper is organized as follows: in Section~\ref{sec:mocks} I describe the mock datasets used to derive eigenbases and test the denoising model. I present the multi-eigenbasis denoising methodology in Section~\ref{sec:method}. In Section~\ref{sec:results} I demonstrate the performance of the model using matrix-level reconstruction metrics and by sampling the posteriors of cosmological parameters. Finally, I summarize the methodology and discuss caveats and future work in Section~\ref{sec:discussion}.

\section{Synthetic Data}
\label{sec:mocks}
In this section I describe the synthetic (mock) data used to develop the denoising methodology. Section~\ref{subsec:desimocks} describes the mocks themselves, focusing on the two suites of DESI DR1 mocks used in this work. In Section~\ref{subsec:covestimation} I describe the standard correlation function and covariance matrix estimation procedures employed in the DESI \lya forest analysis.

\subsection{DESI DR1 Mocks}
\label{subsec:desimocks}
In this work, I use a total of $300$ synthetic realizations of the first year of the DESI \lya forest survey. Below I describe two different suites of mocks available for the DR1 \lya forest dataset: Ly$\alpha$CoLoRe and Saclay mocks. Both mock suites are built from large-scale Gaussian random density fields and use a log-normal transformation to model the \lya forest opacity, but differ in their treatment of redshift-space distortions and the coupling between the quasar and \lya forest density fields, as described in Sections~\ref{subsubsec:lyacoloremocks} and \ref{subsubsec:saclaymocks}. \cite{cuceu_desi_lya_validation} found that the Saclay mocks provide a better fit to the data cross-correlation, while the Ly$\alpha$CoLoRe mocks provide a better fit to the data auto-correlation.

The products of the \texttt{LyaCoLoRe}\footnote{\url{https://github.com/igmhub/LyaCoLoRe}} and \texttt{Saclay} software packages are raw flux transmission files. Realistic quasar spectra are then generated by \texttt{quickquasars}\footnote{\url{https://github.com/desihub/desisim/blob/main/py/desisim/scripts/quickquasars.py}}, which multiplies the flux transmission fields by an unabsorbed continuum template \citep{herrera-alcantar_synthetic_2023}. This continuum template is produced by \texttt{simqso}\footnote{\url{https://github.com/imcgreer/simqso}}~\citep{simqso}, a tool for simulating quasar spectra and photometry. The continuum is modeled as a broken power-law and is tuned to better match the mean quasar continuum measured in eBOSS DR16 \citep{bourboux_completed_2020} using principal component analysis. Each mock is designed to be representative of the DESI DR1 \lya forest dataset, with instrument noise, footprint, and quasar redshift and magnitude distributions tuned to the observed sample. The mocks I use in this work also contain contaminants including quasar redshift errors and absorption due to metals, broad absorption line (BAL) systems, and high column density systems (HCDs) in order to better match the properties of the real data. \cite{herrera-alcantar_synthetic_2023} provides a detailed description of how these contaminants are incorporated. To match the approach used in the baseline DESI \lya analysis \citep[see e.g.,][]{desi_y1_4,desi_dr2_lya,cuceu_2025}, I use a fitted continuum in each forest to measure the flux transmission field rather than the true unabsorbed continuum. Of the $300$ total mock realizations, $100$ Ly$\alpha$CoLoRe and $50$ Saclay mocks correspond to those used in \cite{cuceu_desi_lya_validation}. The remaining $150$ were generated from the same large-scale density fields but with inverted random seeds for quasar down-sampling, resulting in only a small fraction of quasars in common with their paired counterparts.

\subsubsection{Ly$\alpha$CoLoRe Mocks}
\label{subsubsec:lyacoloremocks}
The majority of the mocks are drawn from the Ly$\alpha$CoLoRe suite, comprising $200$ realizations. These mocks are generated in two main stages, which are described in detail in \cite{farr_lyacolore_2020}. First, the \texttt{CoLoRe}\footnote{\url{https://github.com/damonge/CoLoRe}} \citep{ramirez-perez_colore_2022} code generates a large-scale Gaussian random density field given an input matter power spectrum and places quasars in over-dense regions following a prescribed number density and bias. Line-of-sight density and velocity skewers are extracted along each quasar sightline. Second, the \texttt{LyaCoLoRe} code post-processes these skewers to produce realistic \lya forest spectra: small-scale power is added to match the observed one-dimensional forest power spectrum, a log-normal transformation converts the Gaussian field to a physical density field, and the fluctuating Gunn-Peterson approximation (FGPA) is used to compute the \lya optical depth $\tau$. Redshift-space distortions are applied by shifting the optical depth field according to the line-of-sight peculiar velocities computed from the gradient of the Newtonian gravitational potential, and finally the transmitted flux fraction is computed as $F = e^{-\tau}$. The final transmission files are used as input to \texttt{quickquasars} to generate realistic quasar spectra, as described above.

\subsubsection{Saclay Mocks}
\label{subsubsec:saclaymocks}
The remaining $100$ mock realizations are drawn from the Saclay suite, detailed in \cite{etourneau_saclay_2024}. Like the Ly$\alpha$CoLoRe mocks, these begin from Gaussian random fields and apply a log-normal transformation together with the FGPA to produce \lya forest transmission skewers. However, the Saclay mocks differ from Ly$\alpha$CoLoRe in two key aspects. Rather than using a single matter density field for both the \lya forest and quasar positions, the Saclay pipeline generates two independent fields with different input power spectra, each calibrated so that the resulting log-normal fields have the correct linear power spectrum for their respective tracer. Additionally, redshift-space distortions are applied via the velocity gradient along each line of sight using a modified form of the FGPA, rather than by shifting the optical depth field based on the Newtonian gravitational potential. Similar to the Ly$\alpha$CoLoRe case described above, these raw transmission files are then used as input to \texttt{quickquasars}.

\subsection{Correlation Function and Covariance Estimation}
\label{subsec:covestimation}
In the \lya forest 3D analysis, the auto-correlation of the forest flux transmission field and its cross-correlation with quasars are measured in bins of comoving separation along and across the line-of-sight \citep[e.g.,][]{bourboux_completed_2020,desi_y1_4,desi_dr2_lya,cuceu_2025}. The auto-correlation is symmetric along the line-of-sight and therefore measured from separations of $0<r_{\parallel,\perp}<200\,h^{-1}\,\mathrm{Mpc}$ in each direction. The cross-correlation is not symmetric along the line-of-sight since the \lya forest pixel may lie either in front of or behind its neighboring quasar. The cross-correlation is therefore measured over $-200<r_\parallel<200\,h^{-1}\,\mathrm{Mpc}$ along the line of sight and $0<r_\perp<200\,h^{-1}\,\mathrm{Mpc}$ in the transverse direction. With a typical bin size of $\Delta r=4\,h^{-1}\,\mathrm{Mpc}$, there are $50\times50=2500$ bins in the auto-correlation and $50\times100=5000$ bins in the cross-correlation. In the baseline DESI analysis \citep{desi_y1_4,desi_dr2_lya,cuceu_2025}, separate auto- and cross-correlation functions are measured using the \lya absorption in the \lya (or $A$) region ($1040-1205$~\AA~in the rest frame) as well as \lya absorption in the Ly$\beta$ (or $B$) region ($920-1020$~\AA~in the rest frame), yielding four correlation function measurements in total. In this work I only use the $A$ region, both for simplicity and to limit the size of the data vector given the limited number of available DR1 mock realizations. For all cosmological inference, I fit the correlation functions over isotropic scale ranges of $30<r<180\,h^{-1}\,\mathrm{Mpc}$ for the auto-correlation and $40<r<180\,h^{-1}\,\mathrm{Mpc}$ for the cross-correlation.

\begin{figure*}[!t]
    \centering
    \includegraphics[
        width=\textwidth,
        height=0.8\textheight,
        keepaspectratio
    ]{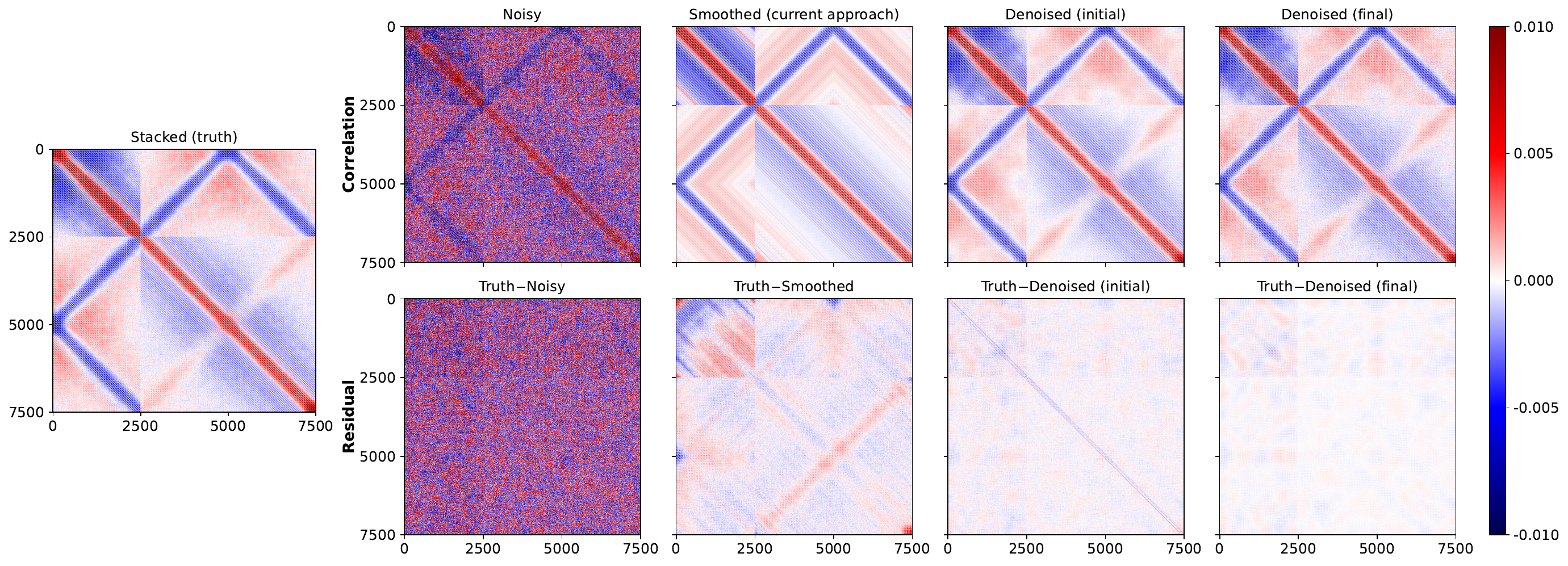}
    \caption{Several correlation and residual matrices for visualization purposes. The correlation matrix on the left is computed from a stack of $\sim100{,}000$ correlation function measurements from Saclay mocks and is taken as the ``truth" for this suite. \textit{Top row, left to right:} the correlation matrix measured from $\sim1000$ sub-samples of a single mock (representative of a survey
    measurement); the smoothed version produced by the current DESI \lya forest
    analysis procedure; the initial denoised estimate from this work; and the final denoised estimate from this work. \textit{Bottom row:} residuals of each matrix with respect to the truth. Notably, the current smoothing approach does not capture all the structure in the correlation matrix, while the final denoised approach presented in this work reconstructs these features more effectively. In all panels, axes correspond to bins in ($r_\parallel,r_\perp$) separation.}
    \label{fig:correlations}
\end{figure*}

The \texttt{picca} package is the DESI \lya forest analysis pipeline software, which fits a continuum to each forest, computes the flux transmission field, measures correlation functions and power spectra, and computes covariance matrices \citep{bourboux_completed_2020}. The covariance matrix is estimated via sub-sampling of correlation function measurements $\xi$ across $\sim1000$ \texttt{HEALPix} pixels \citep{healpix_2005} with \texttt{NSIDE = 16}. For two bins $M$ and $N$ of the correlation function, the covariance matrix is
\begin{equation}
    C_{MN} = \frac{1}{W_MW_N} \sum_{ss} W_M^{ss} W_N^{ss} [\xi_M^{ss}\xi_N^{ss}-\xi_M\xi_N],
    \label{eq:covariance}
\end{equation}
where $ss$ denotes a sub-sample, $W_M^{ss}$ and $\xi_M^{ss}$ are its summed weight and correlation function, respectively, and $W_M = \sum_{ss} W_M^{ss}$ \citep{bourboux_completed_2020}. Small correlations between sub-samples are ignored in this expression. One \texttt{NSIDE = 16} \texttt{HEALPix} pixel subtends an area of $\sim(250\,  h^{-1} \rm Mpc)^2$ at the effective redshift of $z_\mathrm{eff}=2.33$, so the assumption of sub-sample independence for correlations measured up to separations of $r<200\,h^{-1}\,\mathrm{Mpc}$ is justified, especially considering that the current continuum fitting approach suppresses power on large scales \citep[see e.g.,][]{busca_distortion_matrix,turner_2026}. In addition to computing the covariance matrix for the auto- and cross-correlation functions, \texttt{picca} also computes the cross-covariance between these statistics. When using only the $A$ region as in this work, this yields a $7500\times7500$ covariance matrix, for which a reliable noiseless estimate can be obtained by stacking many mocks. For the Ly$\alpha$CoLoRe suite I stack all $200$ mocks ($\sim200{,}000$ sub-samples), and for the Saclay suite I stack all $100$ mocks ($\sim100{,}000$ sub-samples). Even in the Saclay case, the ratio of the number of samples $N$ to the size of the unmasked data vector $n$ is $N/n>10$ (rising to $N/n>20$ when the data vector is masked to the standard fitting range), which should be sufficient for determining the true covariance. The leftmost panel of Figure~\ref{fig:correlations} shows the true correlation matrix for the Saclay mock suite.

The standard estimators of the correlation function and its associated covariance used in the DESI \lya forest analysis are sub-optimal. This is because all pixels in the forest are assumed to be independent, each receiving independent pixel weights during correlation function estimation. In contrast, an optimal estimator would take into account the full covariance of all pixels in the survey, including the covariance between pixels in the same forest and between different forests \citep[e.g.,][]{slosar_2013_lya_boss,ramirez_lya_edr}. Such an optimal estimator would be computationally prohibitive for a survey of this scale. Instead, \cite{ramirez_lya_edr} introduced an improved weighting scheme calibrated on mock datasets that up-weights the large-scale structure contribution to each pixel's variance and improves the precision of the correlation function measurement. The sub-sampling procedure to estimate the covariance matrix is sub-optimal as well, as an optimal estimator would compute the mock-to-mock covariance without reliance on sub-samples, requiring either an infeasibly large number of mocks or compression of the data vector \citep[e.g.,][]{philcox_2021,gerardi_2024_compression, karacayli_2026_multipoles}. The current estimators have been validated in the context of the standard \lya forest analysis \citep[e.g.,][]{cuceu_desi_lya_validation,casas_2025_validation}. The development of a more optimal estimator of the correlation function and its covariance are important future directions, but are beyond the scope of this work. The goal of this work is instead to develop an empirically-motivated methodology to denoise the data-driven covariance matrix, conditioned on mocks.

While the \lya covariance can be well-measured from a stack of several mocks, estimating it from data alone is considerably more difficult. In practice, the covariance matrix of the data can only be measured through sub-sampling, resulting in a very noisy estimate. The first figure in the upper panel of Figure~\ref{fig:correlations} illustrates this by showing the correlation matrix estimated from a single Saclay mock. In this noisy regime, $N<<n$, rendering the sample covariance non-invertible. The standard remedy is to apply a smoothing algorithm to the off-diagonal elements of the correlation matrix $R$, defined by
\begin{equation}
    R_{MN} \equiv \frac{C_{MN}}{\sqrt{C_{MM}C_{NN}}},
    \label{eq:corrmatrix}
\end{equation}
where $C_{MM}$ ($C_{NN}$) are the variances in bins $M$ ($N$). During smoothing, all off-diagonal elements of the correlation matrix with equal separations in $|r_\parallel(M)-r_\parallel(N)|$ and $|r_\perp(M)-r_\perp(N)|$ are replaced with their mean \cite[see e.g.,][]{delubac_boss_dr11, bautista_2017, bourboux_completed_2020, cuceu_desi_lya_validation, desi_y1_4, desi_dr2_lya}. This procedure has been validated for BAO and AP parameter recovery in the standard \lya forest analysis \citep{cuceu23b,cuceu_desi_lya_validation,cuceu_2025}, though recent work suggests that it may result in slightly underestimated uncertainties on the AP scale parameter \citep{turner_2026}. Beyond this, the smoothing procedure imposes a strong structural assumption on the covariance, which causes it to erase genuine large-scale correlation structure. The second panel of the top row in Figure~\ref{fig:correlations} shows the smoothed version of the noisy matrix to its left. Compared to the truth, the smoothed matrix clearly suppresses certain features (e.g., due to cosmic variance). This motivates the development of a more principled approach to denoise the non-invertible sample covariance matrix.

\section{Methodology}
\label{sec:method}
In this section I present the denoising methodology. The methodology is reliant on access to multiple mock suites from which noiseless (``true") covariance matrices can be measured by stacking many sub-samples across realizations. In brief, the steps are as follows:
\begin{enumerate}
    \item Derive a reference correlation matrix from mocks and compute its eigenvectors.
    \item For any given noisy input covariance matrix, project its normalized correlation matrix onto the reference eigenbasis.
    \item Use the pseudo-eigenvalues (diagonal elements of the projected matrix) to reconstruct an initial denoised estimate in its original basis (Sec.~\ref{subsec:stage1}).
    \item Classify the noisy correlation matrix by its pseudo-eigenvalues to determine a weighted residual reference eigenbasis (Sec.~\ref{subsec:stage2}).
    \item Project the noisy residual onto this new eigenbasis, and repeat step 3 above to obtain a denoised residual correction. Add this to the initial estimate to produce the final denoised correlation matrix, then rescale the variance to match that of the input covariance.
\end{enumerate}
I elaborate on each step in the following subsections.

\subsection{Initial Reconstruction}
\label{subsec:stage1}
The first stage of the denoising algorithm begins by constructing a reference correlation matrix from mocks. I use $100$ Saclay and $200$ Ly$\alpha$CoLoRe mocks, each containing $\sim1000$ \texttt{HEALPix} pixel sub-samples, as these represent the full set of available DESI DR1 \lya mocks. For each mock suite, I compute the noiseless (true) covariance matrix from the stack of $N_\mathrm{mocks}\times N_\mathrm{HEALPix}$ correlation functions and rescale it to the single-mock level by multiplying by a factor of $N_\mathrm{mocks}$. I then take the unweighted mean of these two true correlation matrices as the reference matrix.

Decomposing the reference correlation matrix gives
\begin{equation}
    R_{\rm ref} = V \Lambda V^T,
    \label{eq:ref_decomp}
\end{equation}
where $V$ is an orthogonal matrix of eigenvectors and $\Lambda$ is a diagonal matrix containing the eigenvalues of $R_\mathrm{ref}$. The noisy correlation matrix can then be projected onto the basis spanned by $V$,
\begin{equation}
    \tilde{R} = V^T R_{\rm noisy} V.
    \label{eq:projection}
\end{equation}
The projected matrix $\tilde{R}$ is approximately diagonal: the pseudo-eigenvalues $\tilde\lambda_{k}$ along its diagonal represent the contribution of each reference eigenvector to the noisy matrix, while the off-diagonal elements consist primarily of noise. I verify this assumption by projecting the true correlation matrices from each mock suite, $R_{s,\mathrm{truth}}$, and confirming that the result is well approximated by a diagonal matrix (see Appendix~\ref{ap:ref_basis}).

I reconstruct an initial denoised estimate in the original basis, $\hat{R}^\mathrm{initial}$, by retaining only the pseudo-eigenvalues when rotating back from the reference basis: 
\begin{equation}
    \hat{R}^\mathrm{initial} = V\,\mathrm{diag}(\tilde{\lambda}_1, \ldots, \tilde{\lambda}_n)\,V^T.
    \label{eq:initial_recon}
\end{equation}

\subsection{Residual Correction}
\label{subsec:stage2}

\subsubsection{Covariance Classification}
\label{subsec:classification}
The off-diagonal elements of the projected matrix in Equation~\ref{eq:projection} consist of a combination of noise and a smaller signal component arising from the fact that the eigenbasis of the reference correlation matrix (Eq.~\ref{eq:ref_decomp}) does not fully capture the signal in a different noisy correlation matrix. This can be understood by noting that the residual between the true correlation matrix of any mock suite $s$ and its initial reconstruction,
\begin{equation}
    \Delta_s = R_{\rm truth}^{s} - \hat{R}_{\rm truth}^{\mathrm{initial},s},
    \label{eq:clean_resid}
\end{equation}
retains a small amount of structure not captured by the initial reconstruction. To correct for this, I perform a second projection and reconstruction using an eigenbasis derived from a weighted mean of noiseless residuals across mock suites, where the weights are determined by classifying the noisy input matrix. 

\begin{figure*}[!t]
    \centering
    \includegraphics[width=0.8\textwidth]{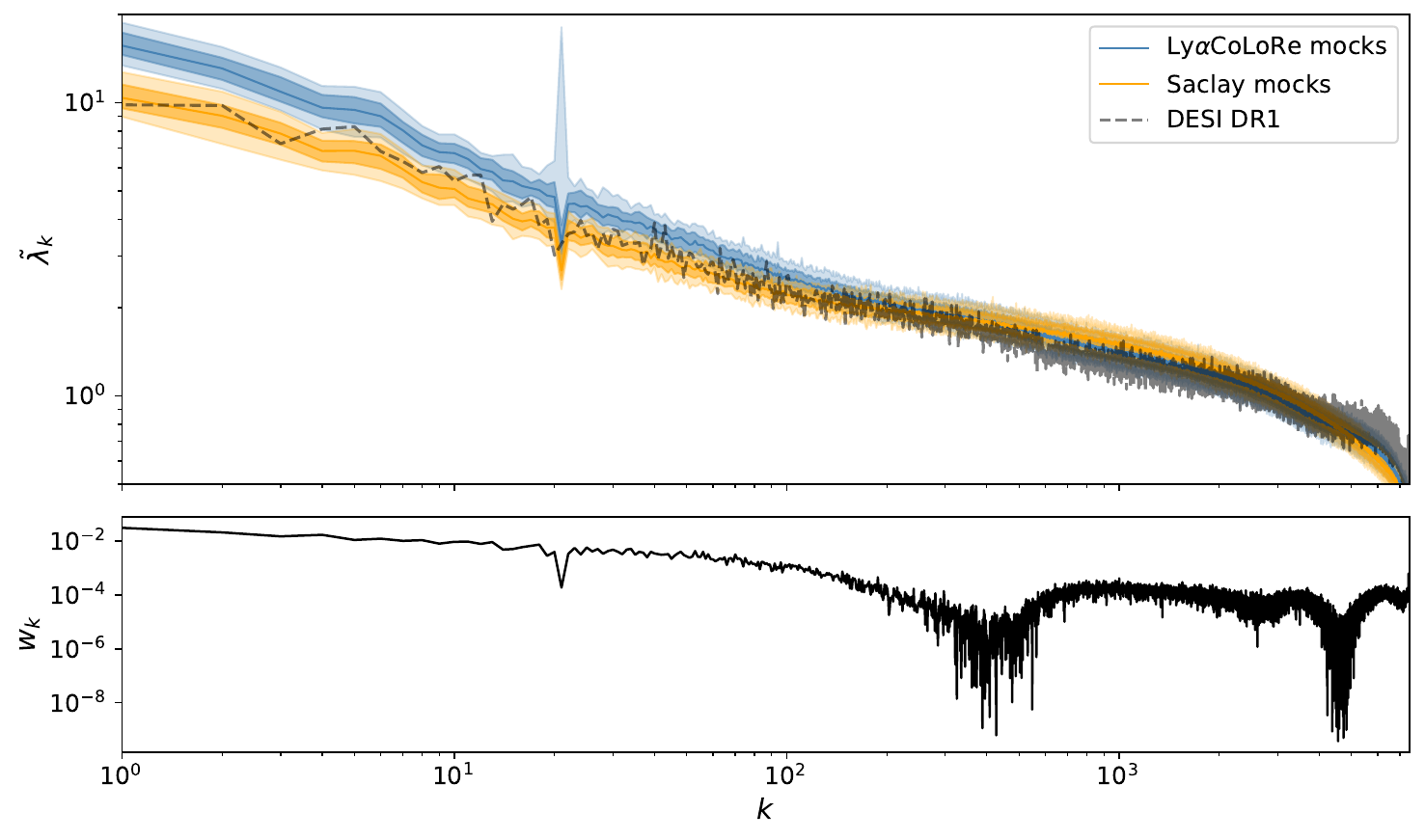}
    \caption{Pseudo-eigenvalue spectra and classifier feature weights. \textit{Top:} pseudo-eigenvalues $\tilde\lambda_k$ as a function of eigenmode index $k$. Solid lines show the median and shaded regions show the 16th--84th and 2.5th--97.5th percentile intervals across the classifier training set for the Ly$\alpha$CoLoRe (\textit{blue}) and Saclay (\textit{orange}) mocks. The pseudo-eigenvalues of the DESI DR1 Ly$\alpha$ forest covariance (\textit{gray dashed line}) are consistent with the mock distributions, validating the applicability of the reference eigenbasis to real data. \textit{Bottom:} feature weights $w_k$ (Eq.~\ref{eq:feature_weights}) used in the classifier model. Weights are largest at low $k$, where the pseudo-eigenvalues are largest in magnitude and the between-to-within suite variance ratio $r_k$ (Eq.~\ref{eq:var_ratio}) is most significant, and remain non-negligible at intermediate and large $k$ where $r_k$ continues to provide discriminating power. Classification of the DESI DR1 covariance provides mock suite weights (Eq.~\ref{eq:suite_weights}) of ($p_\mathrm{Saclay}, p_\mathrm{Ly\alpha CoLoRe}$) $\approx$ ($0.32, 0.68$).}
    \label{fig:eigenspectra}
\end{figure*}

Out of $300$ noisy correlation matrices available, I set aside $20\%$ from each mock suite ($20$ for Saclay, $40$ for Ly$\alpha$CoLoRe) for testing purposes and use the remaining $80\%$ to fit a classifier model. The classifier operates on the pseudo-eigenvalues of the noisy input matrix, computing its distance to each mock suite, $d_s$, as
\begin{equation}
    d_s = \sqrt{\sum_{k=1}^{n} w_k 
    \left(\frac{\tilde{\lambda}_k - \mu_{s,k}}{\sigma_{s,k}}\right)^2},
    \label{eq:distance}
\end{equation}
where $\tilde\lambda_k$ is the vector of pseudo-eigenvalues with $k$ modes, and $\mu_{s,k}$ and $\sigma_{s,k}$ are the mean and standard deviation of pseudo-eigenvalues in the training set for suite $s$ at mode $k$, respectively. The distance to each suite also depends on feature weights $w_k$, a vector over eigenmode index $k$ computed from the training set as
\begin{equation}
w_k = \frac{r_k \, |\bar{\tilde{\lambda}}_k|}{\sum_{k'=1}^n r_{k'} \, |\bar{\tilde{\lambda}}_{k'}|},
\label{eq:feature_weights}
\end{equation}
where
\begin{equation}
    r_k = \frac{\sigma^2_{\mathrm{between},k}}{\sigma^2_{\mathrm{within},k}}
    \label{eq:var_ratio}
\end{equation}
is the between-to-within suite variance ratio for mode $k$. The weight of each mode $k$ is therefore determined by both the mean magnitude of its pseudo-eigenvalue among the training set, $|\bar{\tilde{\lambda}}_k|$, and the between-suite to within-suite variance ratio for that mode, $r_k$. This means that modes where mock suites can be more easily distinguished are assigned more weight during classification. Figure~\ref{fig:eigenspectra} shows the pseudo-eigenvalue distributions for each suite in the training set along with the feature weights. Also shown are the pseudo-eigenvalues of the DESI DR1 \lya covariance matrix. The data pseudo-eigenvalues fall into the distributions spanned by the mocks across most modes, demonstrating the applicability of the reference eigenbasis to real data.

Converting the distances $d_s$ (Eq.~\ref{eq:distance}) to per-suite weights via inverse distance-squared weighting gives
\begin{equation}
    p_s = \frac{d_s^{-2}}{\displaystyle\sum_{s'=1}^{N_s} d_{s'}^{-2}},
    \label{eq:suite_weights}
\end{equation}
where $N_s=2$ is the number of mock suites. When $p_s$ is converted to a hard classification, the classification model achieves $100\%$ accuracy on both the training and testing sets. The weighted mean reference residual matrix is then
\begin{equation}
    \bar{\Delta} = \sum_{s=1}^{N_s} p_s \Delta_s,
    \label{eq:weighted_resid}
\end{equation}
where $\Delta_s$ is the true residual matrix for mock suite $s$ (Eq.~\ref{eq:clean_resid}) and $\sum_{s=1}^{N_s} p_s = 1$.

\subsubsection{Weighted Residual Reconstruction}
\label{subsec:residualcorrection}

To implement the residual correction, I first eigendecompose the reference residual matrix computed in Eq.~\ref{eq:weighted_resid},
\begin{equation}
    \bar{\Delta} = U \Sigma U^T,
    \label{eq:resid_decomp}
\end{equation}
where $U$ is an orthogonal matrix containing the eigenvectors of $\bar{\Delta}$ and $\Sigma$ is a diagonal matrix of its eigenvalues. I compute the noisy residual and project it onto the eigenbasis spanned by $U$,
\begin{equation}
    \Delta_{\rm noisy} = R_{\rm noisy} - \hat{R}^{\mathrm{initial}}, \qquad
    \tilde{\Delta} = U^T \Delta_{\rm noisy}\, U.
    \label{eq:resid_proj}
\end{equation}
Following the same logic as in Section~\ref{subsec:stage1}, I assume the pseudo-eigenvalues $\tilde{\sigma}_k$ of the projected residual capture the missing signal, and reconstruct a denoised residual estimate as
\begin{equation}
    \hat{\Delta} = U\,\mathrm{diag}(\tilde{\sigma}_1, \ldots, \tilde{\sigma}_n)\,U^T.
    \label{eq:resid_recon}
\end{equation}
Adding this to the initial reconstruction (Eq.~\ref{eq:initial_recon}) gives the final denoised correlation matrix,
\begin{equation}
    \hat{R}^{\mathrm{final}} = \hat{R}^{\mathrm{initial}} + \hat{\Delta}
    \label{eq:final_recon}.
\end{equation}
Finally, I convert the denoised correlation matrix back to a covariance matrix, scaling by the input variance. This is an important step as the variance estimated from mock realizations is known to underestimate the variance computed from the data \citep[e.g.,][]{karacayli_2026_multipoles}.

\section{Results}
\label{sec:results}
In this section I present the main results. In Section~\ref{subsec:metrics} I evaluate the model's performance using several matrix-level reconstruction metrics that quantify the quality of the denoised covariance relative to the noiseless covariance estimated from a stack of mocks, which is taken as the truth. In Section~\ref{subsec:cosmoparams} I compare the effects of the denoised covariance, the true covariance, and the current smoothing approach on cosmological parameter estimation.

\subsection{Covariance Reconstruction Metrics}
\label{subsec:metrics}
I assess the reconstruction quality of the covariance matrix using three metrics. The first is the Frobenius norm of the correlation matrix relative to its truth,
\begin{equation}
    \lVert R_\mathrm{truth}-\hat{R} \rVert_F = \sqrt{\sum_{i,j} \left(R_{ij,\mathrm{truth}} - \hat{R}_{ij}\right)^2},
    \label{eq:frob_norm}
\end{equation}
where $R_\mathrm{truth}$ is the true noiseless correlation matrix for the mock suite in question and $\hat{R}$ is an estimate of the correlation matrix. Similarly, I also compute the mean squared error (MSE) between the correlation matrix and its truth as
\begin{equation}
    \mathrm{MSE}(R_\mathrm{truth}, \hat{R}) = \frac{1}{n^2} \sum_{i,j} \left(R_{ij,\mathrm{truth}} - \hat{R}_{ij}\right)^2,
    \label{eq:mse}
\end{equation}
where $n=7500$ is the dimension of the matrix.

\begin{figure*}
    \centering
    \includegraphics[
        width=0.7\textwidth
    ]{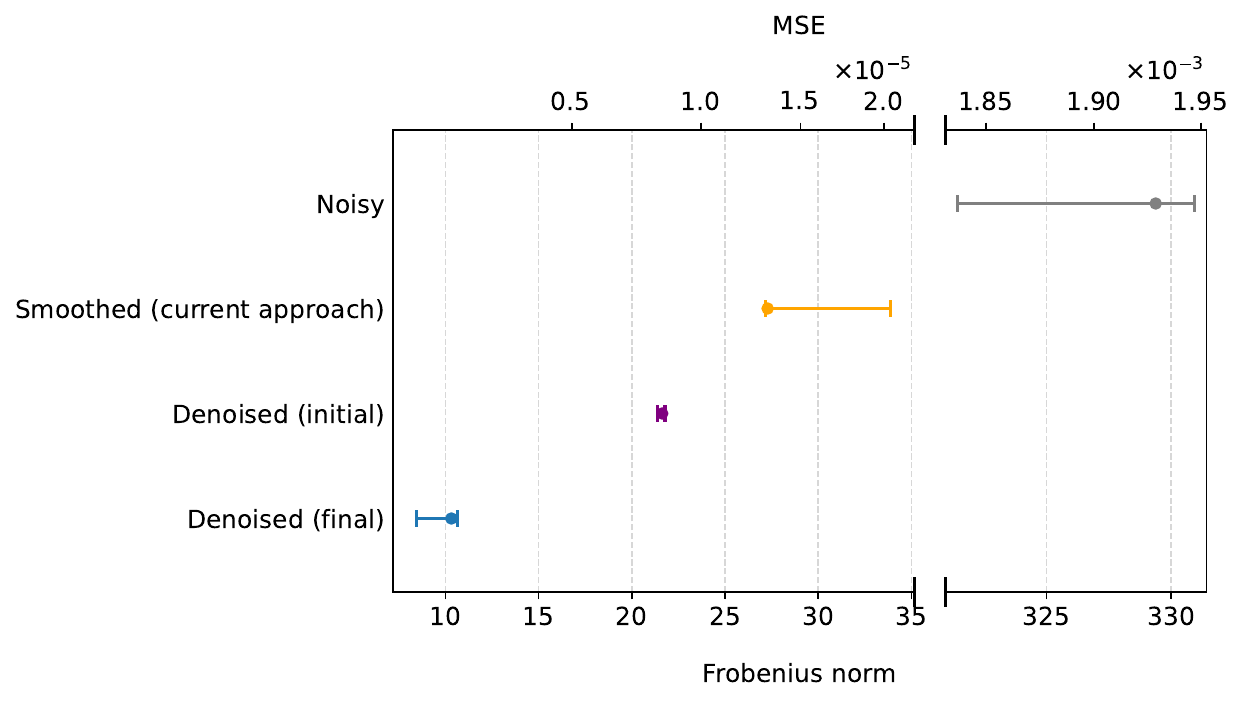}
    \caption{The Frobenius norm (Equation~\ref{eq:frob_norm}, \textit{lower axis}) and MSE (Equation~\ref{eq:mse}, \textit{upper axis}) evaluated on the mock testing set at each pipeline stage relative to the truth: the measured noisy correlation matrix (\textit{gray}), the initial denoised estimate (\textit{purple}), and the final denoised estimate (\textit{blue}). The smoothed matrix from the current approach (\textit{orange}) is also shown for comparison. Points show the median and error bars span the 16th--84th percentile. The Frobenius norm (MSE) drops by a factor of $\sim\!15$ ($\sim\!230$) from the noisy matrix to the initial denoised estimate, and by an additional factor of $\sim\!2$ ($\sim\!4$) from the initial to the final denoised result. The final denoised estimates show larger dispersion than the initial ones, reflecting the added variability from the residual correction stage.}
    \label{fig:stats_progression}
\end{figure*}

Figure~\ref{fig:stats_progression} shows the progression of the Frobenius norm and MSE on the testing set of $60$ mock correlation matrices at each stage of the pipeline: the original noisy matrix, the initial denoised estimate, and the final denoised matrix. The smoothed correlation matrix is also included for comparison. There is a clear improvement in the metrics. The greatest gain comes from the initial reconstruction stage, and the final denoised correlation matrices achieve a median Frobenius norm (MSE) roughly a factor of $\sim30$ ($\sim1000$) smaller than the original noisy matrices.

Lastly, I evaluate the Kullback-Leibler \citep[KL;][]{kullback1951information} divergence $D_\mathrm{KL}$ between two zero-mean multivariate Gaussian distributions with covariances $C_\mathrm{truth}$ and $\hat{C}$, normalized by the matrix dimension,
\begin{equation}
    \begin{aligned}
    \frac{D_{\rm KL}(C_\mathrm{truth} \| \hat{C})}{n} = \frac{1}{2} \Bigg[
    & \frac{1}{n}\mathrm{tr}\!\left(\hat{C}^{-1} C_\mathrm{truth}\right) - 1 \\
    & - \frac{1}{n}\ln \det\!\left(\hat{C}^{-1} C_\mathrm{truth}\right)
    \Bigg],
    \end{aligned}
    \label{eq:kl}
\end{equation}
This quantity measures the loss of information when the true covariance $C_\mathrm{truth}$ is approximated by $\hat{C}$: it is zero when $\hat{C} = C_\mathrm{truth}$ and increases monotonically as $\hat{C}$ departs from the truth. Dividing by $n$ makes the metric comparable across different matrix dimensions. Because the initial noisy covariance is non-invertible, its KL divergence cannot be evaluated and I instead focus on the initial and final denoised covariances.

\begin{figure}
    \centering
    \includegraphics[
        width=\columnwidth
    ]{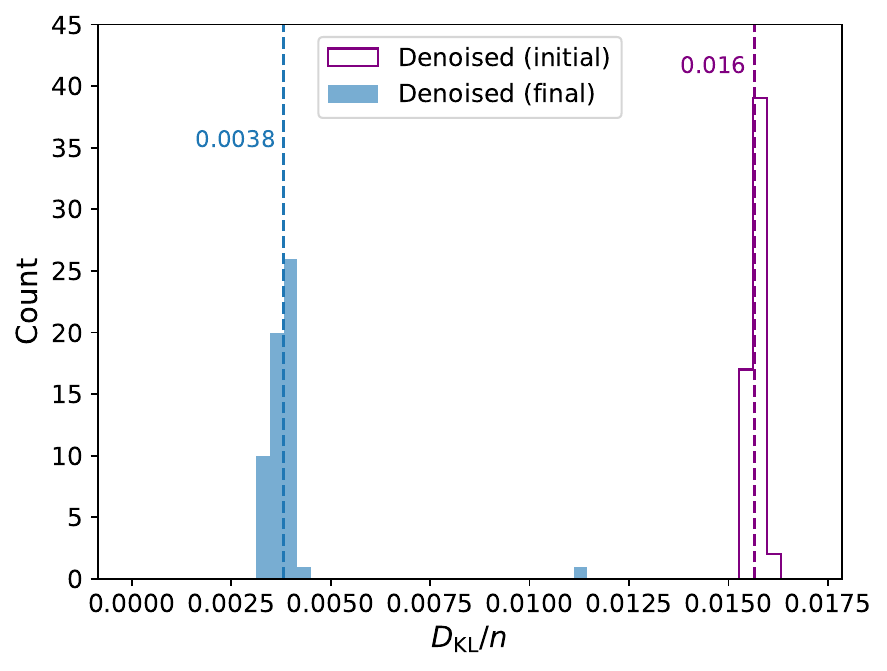}
    \caption{Distributions of the normalized KL divergence $D_\mathrm{KL}/n$ (Eq.~\ref{eq:kl}) for the initial (\textit{purple unfilled}) and final (\textit{blue filled}) denoised covariance matrices in the mock testing set relative to their truth. Vertical dashed lines indicate their respective medians. The median $D_\mathrm{KL}/n$ drops by a factor of $\sim4.2$ when progressing from the initial to final denoised covariances.}
    \label{fig:KL_divergence_progression}
\end{figure}

Figure~\ref{fig:KL_divergence_progression} shows the $D_\mathrm{KL}/n$ distributions on the mock testing set for the initial and final denoised covariance relative to their truth. Two denoised covariances ($\sim3\%$ of the testing set) are not positive definite and excluded from the sample. The median $D_\mathrm{KL}/n$ decreases by a factor of $\sim4.2$ from the initial to the final denoised covariance, consistent with the improvements seen in the Frobenius norm and MSE.

\subsection{Cosmological Parameters}
\label{subsec:cosmoparams}
I perform cosmological inference with the denoised covariances and compare the results against those obtained with the true covariance and current smoothing approach. This is done in the context of a full-shape analysis \citep[e.g.,][]{cuceu_cosmology_2021,cuceu_2025}, which extracts cosmological information from the full shape of the \lya forest two-point correlation functions rather than from the BAO peak position alone. For the \lya forest, a full-shape approach more than doubles the constraining power on Alcock-Paczy\'{n}ski relative to a BAO-only analysis \citep{cuceu_2025}.

Modeling choices for the DESI DR1 mock and data correlations follow those of \cite{cuceu_2025}, with two main differences as mentioned in Section~\ref{subsec:covestimation}: I restrict the analysis to the $A$ region only (typically spanning $1040-1205$~\AA~in the quasar rest frame)\footnote{While $1040-1205$~\AA~is the typical rest-frame wavelength range used for the $A$ region, the fits in this work use the range $1040-1200$~\AA~due to mock analysis availability.}, and fit the auto-correlation over an isotropic scale range of $30<r<180\,h^{-1}\,\mathrm{Mpc}$. For a detailed description of all other modeling choices and the fitting procedures, see \cite{cuceu_2025}.

In the full-shape analysis, cosmological information comes from the BAO scale, the Alcock-Paczy\'{n}ski \citep[AP;][]{AP_1979} effect, and redshift space distortions (RSD). Following \cite{cuceu_2025}, the isotropic BAO parameter is defined as
\begin{equation}
    \alpha_p(z_\mathrm{eff}) = \sqrt{\frac{D_M(z_\mathrm{eff})D_H(z_\mathrm{eff})/r_d^2}{[D_M(z_\mathrm{eff})D_H(z_\mathrm{eff})/r_d^2]_\mathrm{fid}}}
    \label{eq:alpha}
\end{equation}
where the $p$ subscript denotes the BAO peak, $D_M(z)$ is the transverse comoving distance, $D_H(z)=c/H(z)$ with the speed of light $c$ and Hubble parameter $H(z)$, $r_d$ is the acoustic scale, and $z_\mathrm{eff}=2.33$ is the effective redshift of the DESI DR1 \lya dataset. The $\mathrm{fid}$ subscript denotes quantities computed assuming the fiducial cosmology, which is given by \cite{planck_2016} when performing the analysis on DR1 mocks and \cite{planck_2018_cosmology} when performing it on the data. One can also measure the isotropic scale of the broadband, $\alpha_s$, where the $s$ subscript denotes the smooth component. However, this component is more difficult to model cleanly and is typically marginalized over \citep{cuceu_cosmology_2021}; whether it traces cosmology or reflects a combination of cosmology and systematic effects such as continuum fitting distortions remains an open question \citep[see e.g.,][]{busca_distortion_matrix,turner_2026}.

The AP scale parameter from the full shape, including contributions from both the BAO scale and the broadband, is parameterized as
\begin{equation}
    \phi_f(z_\mathrm{eff}) = \frac{D_M(z_\mathrm{eff})H(z_\mathrm{eff})}{[D_M(z_\mathrm{eff})H(z_\mathrm{eff})]_\mathrm{fid}}.
    \label{eq:phi}
\end{equation}
The full-shape analysis is also sensitive to $f\sigma_8$, the product of the cosmic growth rate and the amplitude of matter fluctuations in spheres of $8\,h^{-1}\,\mathrm{Mpc}$. This information comes from the RSD signal recoverable from the joint auto- and cross-correlation analysis, which is described in detail in \cite{cuceu_2025}.

I perform cosmological inference using \texttt{Vega}\footnote{\url{https://github.com/andreicuceu/vega}} with the full shape of the \lya auto- and cross-correlation functions. In brief, a template correlation function built from the fiducial cosmology is fit to the measured correlations using the full covariance matrix including cross-covariance. For all mocks in the testing set, the maximum likelihood solution is found using \texttt{iminuit} and assuming Gaussian posteriors. I perform these fits using the true covariance matrix, the smoothed covariance used in the current approach, the final denoised covariance, and a diagonal covariance for a simple comparison. Additionally, for one mock in the testing set and for the DESI DR1 \lya dataset, I sample the full posteriors using the \texttt{PolyChord} nested sampler. For comprehensive details on the fitting process including nuisance parameters that are marginalized over, see \cite{cuceu_2025}.

\begin{figure*}
    \centering
    \includegraphics[
        width=0.8\textwidth
    ]{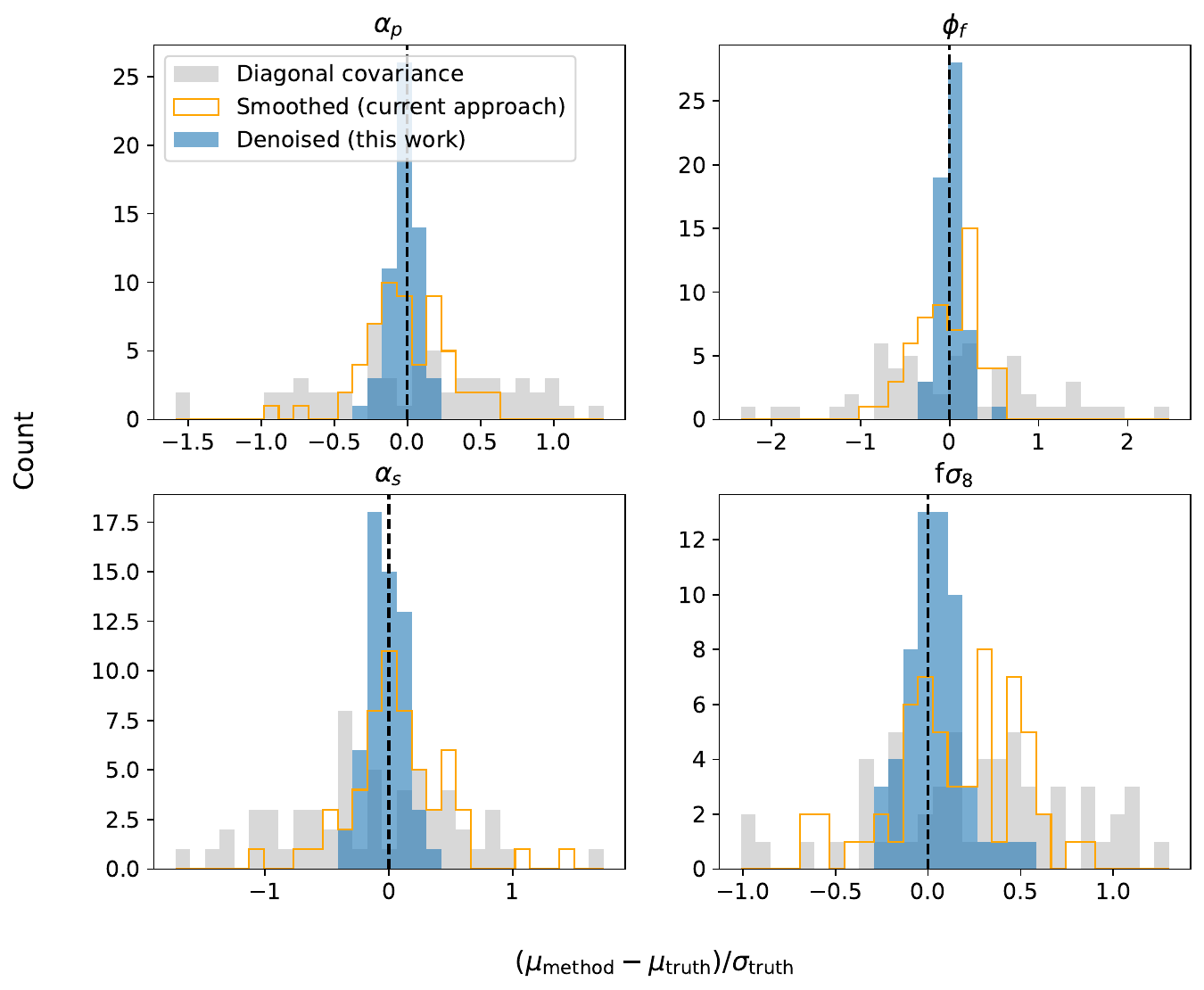}
    \caption{Normalized parameter bias distributions $(\mu_\mathrm{method}-\mu_\mathrm{truth})/\sigma_\mathrm{truth}$ for a selection of parameters: $\alpha_p$ (isotropic BAO scale parameter), $\phi_f$ (full-shape AP scale parameter), $\alpha_s$ (isotropic broadband scale parameter), and $f\sigma_8$. Distributions are computed from maximum likelihood fits on the mock testing set, where $\sigma_\mathrm{truth}$ is the uncertainty inferred from the fit using the true
    covariance and $\mu$ is the mean. A value of zero (\textit{vertical dashed line}) indicates no bias relative to the true covariance fit. I show results using the current smoothing-based covariance (\textit{orange}), the final denoised covariance (\textit{blue filled}), and a diagonal covariance (\textit{gray filled}) for comparison. The denoised covariance produces the narrowest distributions, indicating the most consistent parameter recovery across mocks. No significant systematic bias is found for either the smoothed or denoised covariance. Median and percentile regions for each distribution are reported in Table~\ref{tab:bias_error_combined}.}
    \label{fig:bias_distributions}
\end{figure*}

Figure~\ref{fig:bias_distributions} shows the parameter bias distributions computed from fit results on the mock testing set, normalized by the uncertainty inferred with the true covariance. I find that using a diagonal covariance results in substantial scatter around zero for all parameters, as expected from the loss of information when off-diagonal correlations are ignored. Both the current smoothing approach and the denoised covariance yield distributions centered near zero, indicating no significant systematic bias in parameter recovery for either method. The denoised covariance consistently produces narrower distributions than the smoothing approach, indicating more consistent parameter recovery across mocks. A narrower distribution here is a favorable result, reflecting smaller mock-to-mock scatter in the recovered parameters relative to $\sigma_\mathrm{truth}$. Median and 16th-84th percentile ranges for each bias distribution are reported in Table~\ref{tab:bias_error_combined}.

\begin{figure*}
    \centering
    \includegraphics[
        width=0.8\textwidth
    ]{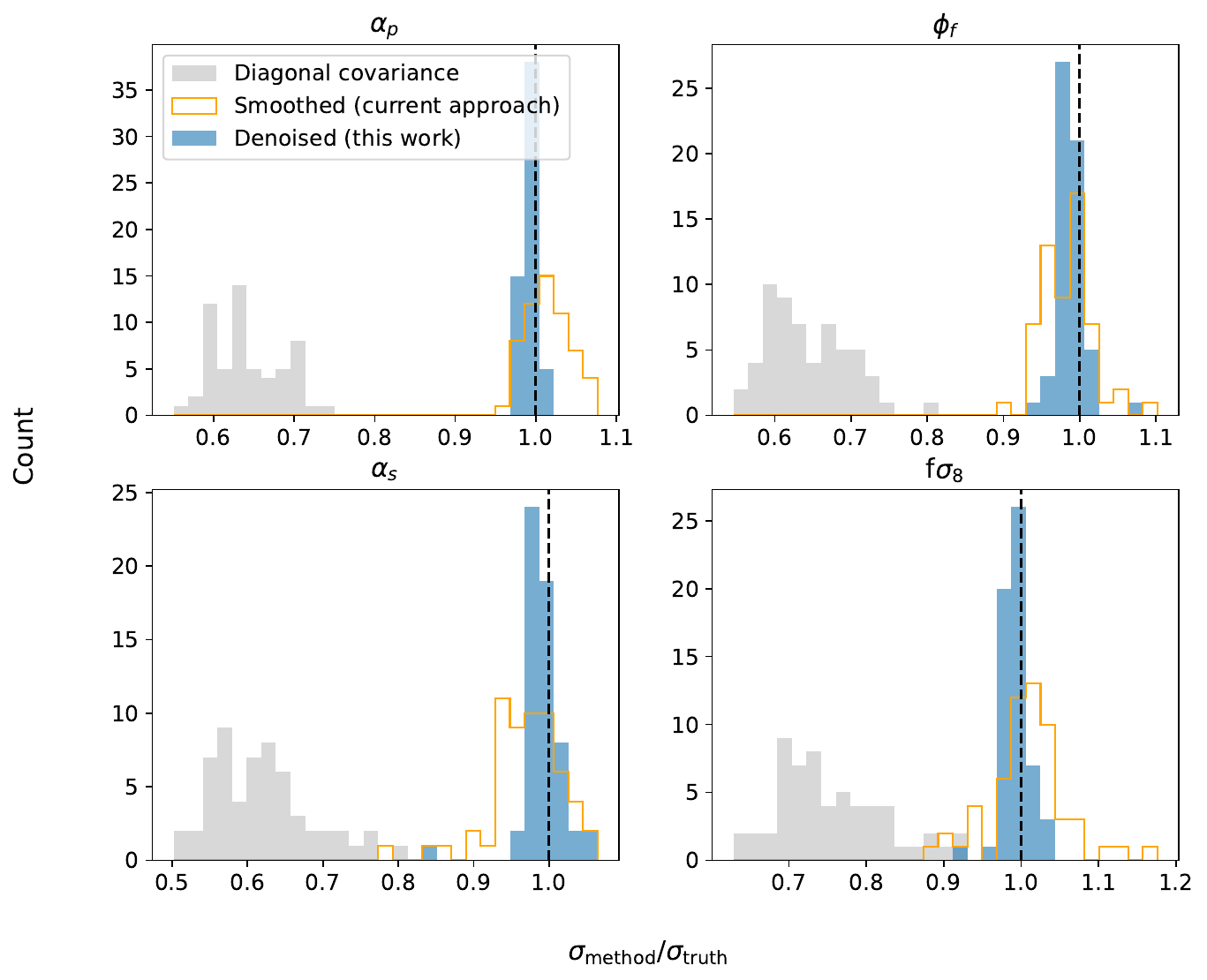}
    \caption{Uncertainty ratio distributions $\sigma_\mathrm{method}/\sigma_\mathrm{truth}$ for the same parameters and covariance methods as in Fig.~\ref{fig:bias_distributions}, computed from fits on the mock testing set. A value of unity (\textit{vertical dashed line}) indicates perfect agreement with the uncertainty inferred from the true covariance fit. The diagonal covariance severely underestimates uncertainties for all parameters shown, while both the smoothing-based (\textit{orange}) and denoised (\textit{blue filled}) covariances yield median ratios near unity. The denoised covariance produces a significantly narrower distribution, indicating more consistent uncertainty recovery across mocks. Median and percentile ranges for each distribution are reported in Table~\ref{tab:bias_error_combined}.}
    \label{fig:sigma_ratios}
\end{figure*}

To assess uncertainty calibration, Figure~\ref{fig:sigma_ratios} shows the uncertainty ratio distributions $\sigma_\mathrm{method}/\sigma_\mathrm{truth}$ across the mock testing set for each covariance method. The diagonal covariance yields severely underestimated uncertainties for all parameters. Both the smoothing-based and denoised covariances produce distributions centered near unity, indicating well-calibrated uncertainties on average. The denoised covariance yields a significantly narrower distribution of ratios more tightly clustered around one, indicating more consistent uncertainty estimation across mocks relative to the smoothing approach. Median values and percentile regions are listed in Table~\ref{tab:bias_error_combined}.

\begin{table*}
    \centering
    \resizebox{0.9\textwidth}{!}{
    \begin{tabular}{l|cc|cc|cc}
    \hline
     & \multicolumn{2}{c}{Denoised (this work)}
     & \multicolumn{2}{c}{Smoothed (current approach)}
     & \multicolumn{2}{c}{Diagonal covariance} \\
    \cline{2-7}
    Parameter
    & Bias & Uncertainty Ratio
    & Bias & Uncertainty Ratio
    & Bias & Uncertainty Ratio \\
    \hline

    $\alpha_p$
    & $-0.018^{+0.092}_{-0.081}$ & $0.989^{+0.011}_{-0.012}$
    & $-0.013^{+0.289}_{-0.256}$ & $1.016^{+0.026}_{-0.027}$
    & $0.104^{+0.630}_{-0.748}$ & $0.634^{+0.064}_{-0.036}$ \\

    $\phi_f$
    & $0.002^{+0.129}_{-0.112}$ & $0.986^{+0.014}_{-0.010}$
    & $-0.003^{+0.303}_{-0.390}$ & $0.984^{+0.027}_{-0.034}$
    & $0.049^{+0.873}_{-0.848}$ & $0.634^{+0.068}_{-0.043}$ \\

    $\alpha_s$
    & $-0.040^{+0.145}_{-0.109}$ & $0.990^{+0.018}_{-0.014}$
    & $0.041^{+0.401}_{-0.287}$ & $0.974^{+0.040}_{-0.039}$
    & $-0.132^{+0.649}_{-0.772}$ & $0.613^{+0.073}_{-0.053}$ \\

    $f\sigma_8$
    & $0.030^{+0.126}_{-0.122}$ & $0.992^{+0.015}_{-0.014}$
    & $0.164^{+0.333}_{-0.296}$ & $1.009^{+0.033}_{-0.029}$
    & $0.321^{+0.513}_{-0.573}$ & $0.738^{+0.082}_{-0.042}$ \\

    \hline
    \end{tabular}
    }
    \caption{Median and 68\% limits of parameter bias (Fig.~\ref{fig:bias_distributions}) and uncertainty ratio (Fig.~\ref{fig:sigma_ratios}) distributions for different covariance methods across the mock testing set.}
    \label{tab:bias_error_combined}
\end{table*}

\begin{figure*}
    \centering
    \includegraphics[
        width=0.8\textwidth
    ]{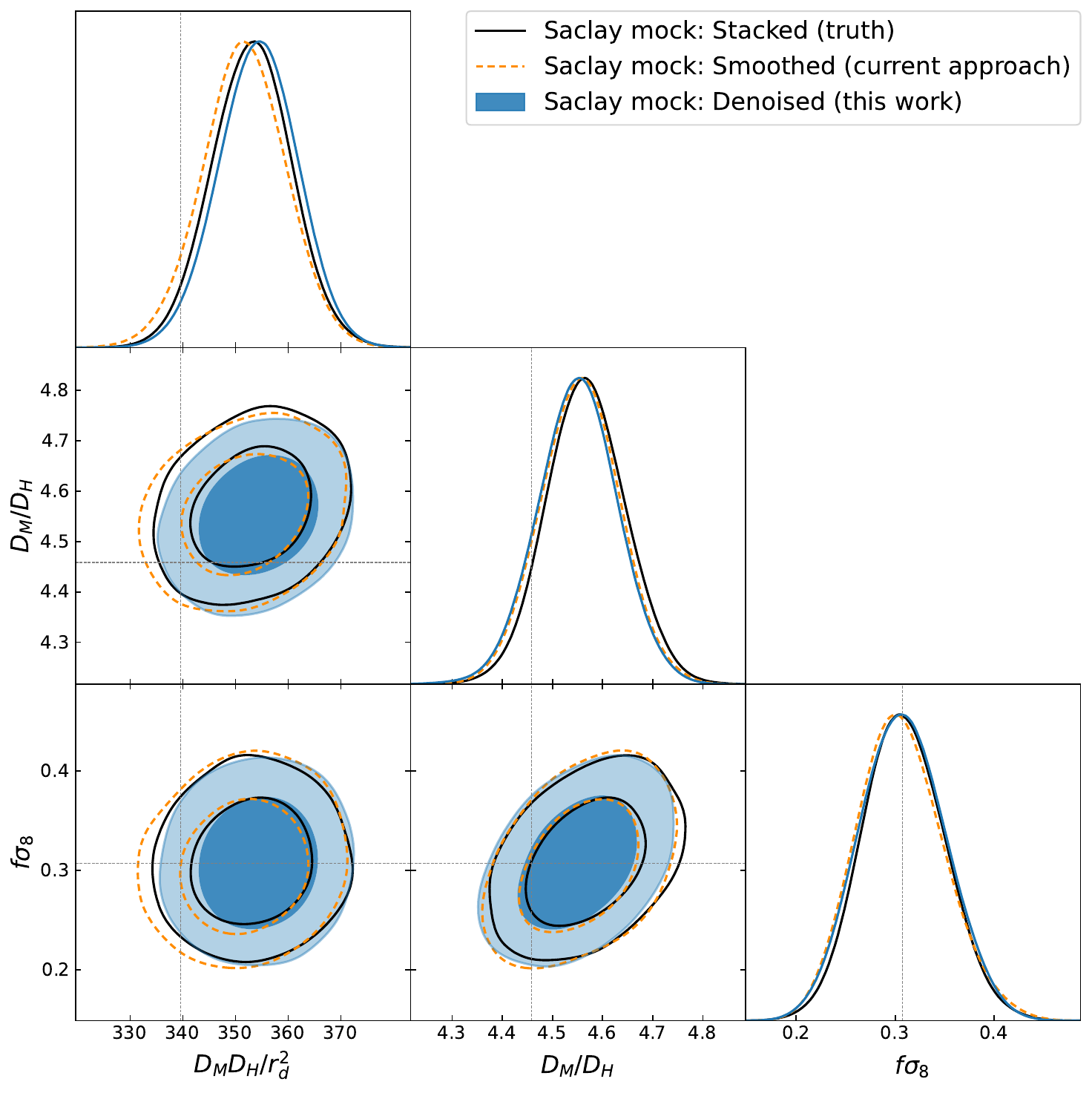}
    \caption{Posteriors for a full-shape analysis on a Saclay mock in the testing set sampled with \texttt{PolyChord}. I compare results using the true covariance measured from a stack of $100$ Saclay mocks with $\sim1000$ sub-samples per mock (\textit{black contours}), the smoothed covariance used in the current approach (\textit{orange dashed contours}), and the final denoised covariance from this work (\textit{blue filled contours}). Cross-hairs indicate the fiducial values used to generate the mock, and all covariance methods recover the fiducial cosmology to within $2\sigma$. The denoised result shows excellent agreement with that of the true covariance.}
    \label{fig:mock_chains}
\end{figure*}

In Figure~\ref{fig:mock_chains} I show the sampled posteriors for one Saclay mock in the classifier testing set. I include results using the true, smoothing-based, and the final denoised covariance, all of which are shown in normalized correlation matrix form in Figure~\ref{fig:correlations}. Here the isotropic BAO and AP scale parameters are converted into cosmological distance quantities via Equations~\ref{eq:alpha} and~\ref{eq:phi}. Across all parameters, there is excellent agreement between the true covariance and denoised covariance results. The smoothed covariance also produces results in agreement for this mock.

\begin{figure*}
    \centering
    \includegraphics[
        width=0.8\textwidth
    ]{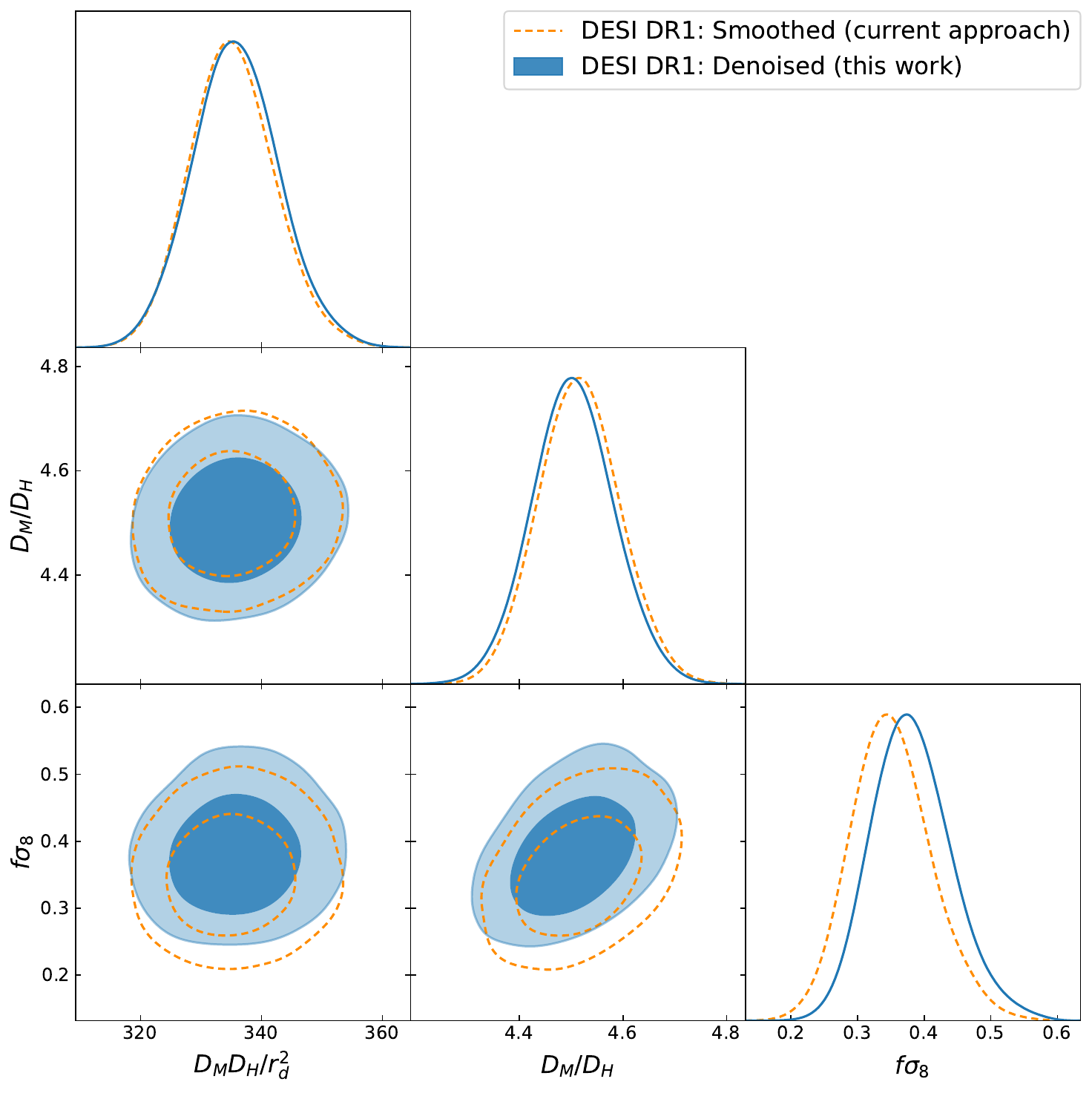}
    \caption{Posteriors for the DESI DR1 \lya full-shape analysis sampled with \texttt{PolyChord} using the standard smoothed covariance (\textit{orange dashed contours}) and the final denoised covariance from this work (\textit{blue filled contours}). The two results are in excellent agreement for quantities derived from the BAO and AP scale parameters. A $\sim 0.5\sigma$ shift in $f\sigma_8$ is present, but this parameter was not demonstrated to be unbiased in \cite{cuceu_2025}.}
    \label{fig:dr1_chains}
\end{figure*}

Finally, Figure~\ref{fig:dr1_chains} shows the posteriors for the DESI DR1 \lya dataset using the smoothed versus denoised covariance. The smoothed result reproduces the DESI DR1 full-shape analysis of \cite{cuceu_2025} up to the analysis differences noted earlier in this section, resulting in larger overall uncertainties. Results for the BAO and AP parameters using the denoised covariance are in excellent agreement with previous results. There is a $\sim 0.5\sigma$ shift in $f\sigma_8$, though this parameter was not shown to be unbiased in the validation analysis on mocks of \cite{cuceu_2025}.

\section{Summary and Discussion}
\label{sec:discussion}
In this work, I developed a multi-eigenbasis approach for denoising data-driven covariance matrices conditioned on mock datasets, using the Ly$\alpha$CoLoRe and Saclay mock suites described in Section~\ref{subsec:desimocks}. The method requires a noiseless mock-based reference covariance matrix that reasonably captures the data covariance. An initial denoised estimate is obtained by projecting the noisy covariance onto the eigenbasis of the reference matrix and reconstructing it using the resulting pseudo-eigenvalues. This estimate is then refined through a weighted residual correction, where the weights are determined by classifying the noisy covariance against at least two available mock suites. I demonstrated that the initial reconstruction alone already outperforms the current smoothing-based approach used in the DESI 3D \lya forest analysis. While developed in the context of this analysis, the methodology is general and could be applied in other contexts, especially where the data vector is very large or the sample covariance is non-invertible.

In the initial reconstruction stage, a noisy correlation matrix is projected onto the reference eigenbasis; the resulting pseudo-eigenvalues encode most of the signal while noise is redistributed into the off-diagonal elements of the projected matrix. These pseudo-eigenvalues serve two purposes: they are used both to construct an initial denoised estimate and to classify the noisy covariance against the available mock suites. The classifier is a weighted distance model trained on $80\%$ of the available mocks, with the remaining $20\%$ reserved for testing purposes. It assigns per-suite weights by converting inverse-squared distances in pseudo-eigenvalue space, which are then used to form a weighted mean reference residual matrix. The noisy residual is projected onto the eigenbasis of this weighted reference residual, and the pseudo-eigenvalues of the projection are used to reconstruct a denoised residual estimate. Adding this residual correction to the initial reconstruction yields the final denoised correlation matrix, which is then rescaled by the measured variance to produce a denoised covariance matrix.

I evaluated the performance of the denoising method using matrix-level reconstruction metrics and \lya full-shape cosmological fits on the mock testing set withheld from the classifier training. I compared results obtained with the denoised covariance against those from the true covariance and the current smoothing approach. The method achieves significant improvements in all matrix-level metrics relative to the current approach. Cosmological parameter values and uncertainties recovered using the denoised covariance are consistent with those from the true covariance, and show substantially more consistent mock-to-mock recovery than the smoothing approach. I also applied the denoised covariance to the DESI DR1 \lya dataset and found consistent cosmological results compared to the current smoothed covariance. The only notable discrepancy is a $\sim0.5\sigma$ shift in $f\sigma_8$; however, this parameter has not been demonstrated to be unbiased in the \lya full-shape analysis \citep{cuceu_2025} and the shift should therefore be interpreted with caution. Finally, I demonstrate that the performance of the denoising approach is not sensitive to the choice of initial reference matrix by repeating the analysis with a Ly$\alpha$CoLoRe-only reference matrix in Appendix~\ref{ap:ref_basis}.

The central assumption underlying this work is that the important features of the data covariance matrix are represented in the mock covariances. I showed that the pseudo-eigenvalues of the data covariance are consistent with the distributions from mocks (Fig.~\ref{fig:eigenspectra}), which addresses this assumption, but a more comprehensive comparison of mock and data covariances may be warranted in future work. This work is also limited by access to only two mock suites, namely the Ly$\alpha$CoLoRe and Saclay mocks. The inclusion of additional and more realistic mock suites in future work should help improve the performance of the denoising method and its applicability to real data, as it will provide greater flexibility to the classification model and potentially capture correlation features not represented by the two suites used in this work.

An important caveat is that while $20\%$ of the mock covariances are withheld from the classifier model training set and reserved for testing purposes, the underlying true covariance of these test-set mocks are still represented in the reference correlation and residual matrices. A more stringent test would withhold an entire mock suite from all reference matrices and the classifier, benchmarking denoiser performance on a suite that played no role in its development. Such a test would be especially informative if the differences between the withheld suite and the training suites are believed to be comparable to or larger than those between the training suites and the data. However, the denoising model requires at least two mock suites in order for the classification to add meaningful information, and only two mock suites are available for the DESI DR1 \lya dataset, making this infeasible for the present work. Future work will re-train and test the model on DESI DR2 or DR3 mocks where additional suites will be available and this type of test becomes possible.

I also note that I did not apply the Hartlap correction \citep{hartlap_2007} to account for the bias in the inverse covariance when the covariance is estimated from a limited number of samples relative to the size of the data vector \citep[see also e.g.,][]{percival_covariance_2014,percival_2022}, nor do I account for the additional parameter uncertainty arising from treating the sample covariance as fixed \citep{dodelson_schneider_2013}, which requires a non-Gaussian likelihood \citep{sellentin_heavens_2016} that can be approximated assuming a Gaussian likelihood with additional correction factors \citep[e.g.,][]{dodelson_schneider_2013,percival_covariance_2014,percival_2022}. I tested that the impact of these corrections is small or negligible in this setting, and stress that the key results of this work lie in the comparison between the true and denoised covariances, for which the absolute correction is likely irrelevant as it applies to both cases. Moreover, deriving the appropriate corrections for the denoised covariance is non-trivial. These correction factors often depend on the number of samples and the data vector dimension, but the effective number of samples for the denoised covariance is a nonlinear function of both the mock-based reference covariance -- which itself is dependent on a different number of samples per mock suite -- and the noisy covariance measured from the data. While the correction factors associated with the mock-based reference are technically non-negligible here, this concern could be mitigated in future work by building the reference from a larger stack of mocks (e.g., $500$ per suite rather than $100-200$).

There are several directions for future work. The most immediate extension is to apply this method to DESI DR2 or DR3 covariances, requiring only a re-computation of reference matrices and re-training of the classifier with the updated mock suites. Access to additional mock suites will improve the denoising model by increasing flexibility in the classification step. A larger number of mocks per suite will also enable the extension of this work to include the $B$ region ($920-1020$~\AA~in the rest frame). The baseline DESI \lya analysis uses both the $A$ and $B$ region to measure correlation functions and constrain cosmological parameters, as discussed in Section~\ref{subsec:covestimation}, doubling the data vector dimension from $7500$ to $15000$. With a sufficiently large stack of mocks, robust noiseless covariance estimates for this larger data vector will be achievable, allowing the denoising methodology to be extended accordingly.

The mocks used in this work were generated assuming $\Lambda$CDM fiducial cosmologies \citep{planck_2016,planck_2018_cosmology}. In light of recent DESI results that hint that dark energy may evolve with redshift \citep[][]{desi_y1_6,desi_dr2_discretetracers}, it will be valuable to test the robustness of this methodology to non-$\Lambda$CDM cosmologies. This will be influenced by the sensitivity of the sample covariance matrix to the choice of fiducial cosmology, which remains an open question.

Another exciting future direction is to extend this work to analyses that use the true (or unabsorbed) continuum of quasars to measure the flux transmission field rather than the current continuum fitting approach mentioned in Section~\ref{subsec:desimocks}. The baseline DESI \lya forest analysis fits a continuum to each forest prior to measuring the flux transmission field, a procedure that suppresses power on large scales by absorbing large-scale density information into the continuum fitting parameters \citep[see e.g.,][]{busca_distortion_matrix,turner_2026}. Several recent efforts have developed methods to estimate the unabsorbed continuum strictly from spectral information outside the forest region \citep[e.g.,][]{suzuki_predicting_2005,paris_principal_2011,liu_deep_2021,turner_2024,pistis_continuum_DL_2025}, which could enable a $\sim10-15\%$ improvement in constraining power on the Alcock-Paczy\'{n}ski parameter \citep{turner_2026} and remove the need for the distortion matrix that complicates the analysis \citep{busca_distortion_matrix}. While the current covariance smoothing procedure has been validated in the context of the standard continuum fitting approach \citep[e.g.,][]{cuceu_desi_lya_validation}, it has not been validated in a true-continuum analysis. Extending the denoising approach developed here to such an analysis is therefore an important future step, enabling future cosmological constraints using estimates of the unabsorbed continuum.

\section*{Acknowledgments}
I thank Andreu Font-Ribera and Paul Martini for useful comments on this manuscript. I also thank Paul Martini, Molly Wolfson, Naim Karaçaylı, and David Weinberg for helpful discussions.
WT acknowledges support from the United States Department of Energy, Office of High Energy Physics under Award Number DE-SC-0011726.

\bibliographystyle{aasjournal}
\bibliography{main}

\begin{appendix}

\section{Sensitivity to Choice of Reference Matrix}
\label{ap:ref_basis}

\begin{figure*}
    \centering
    \includegraphics[
        width=0.7\textwidth
    ]{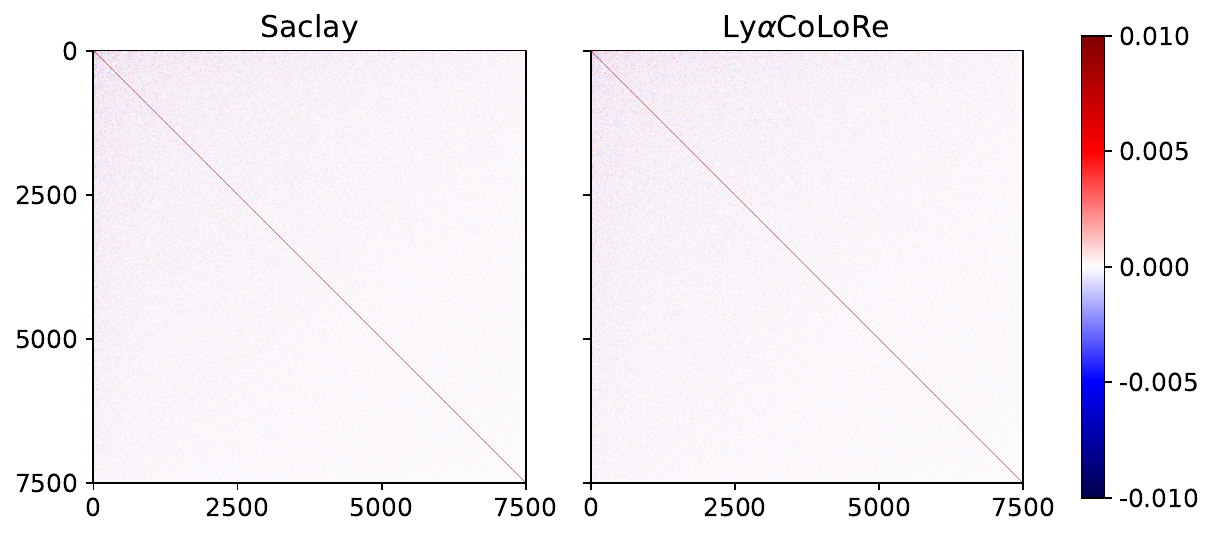}
    \caption{True correlation matrices for each mock suite projected onto the default reference eigenbasis (Eq.~\ref{eq:projection}). \textit{Left:} the projection of the Saclay correlation matrix computed from $100$ mocks and $\sim1000$ sub-samples each. \textit{Right:} the projection of the Ly$\alpha$CoLoRe correlation matrix computed from $200$ mocks and $\sim1000$ sub-samples each. In both cases the diagonal-to-total power ratio (Eq.~\ref{eq:diag_power_ratio}) $\mathcal{P} \approx 0.96$, demonstrating that each matrix is nearly diagonalized in the reference eigenbasis.}
    \label{fig:mock_projections}
\end{figure*}

As discussed in Section~\ref{sec:method}, I derive the initial eigenbasis from a reference correlation matrix based on a combination of Saclay and Ly$\alpha$CoLoRe mocks. I assume that the true correlation matrix of each mock suite is approximately diagonalized in this reference eigenbasis. Figure~\ref{fig:mock_projections} validates this assumption by showing the true correlation matrices of each suite projected onto the default reference eigenbasis. Both projected matrices are very nearly diagonal, with only a small residual signal present in the off-diagonal elements. This can be quantified by computing the ratio of power present in the diagonal to the total power,
\begin{equation}
    \mathcal{P} = \frac{\sum_{k} (V^T R V)_{kk}^2}{\sum_{i,j} (V^T R V)_{ij}^2} = \frac{\sum_{k} \tilde\lambda_k^2}{\|V^T R V\|_F^2},
    \label{eq:diag_power_ratio}
\end{equation}
where $k$ indexes the diagonal elements of the projected matrix, $\tilde\lambda_k$ are the pseudo-eigenvalues, $V$ is the reference eigenbasis, $R$ is the input correlation matrix, and $\|V^T R V\|_F^2$ is the Frobenius norm of the projected correlation matrix. Both the Saclay and Ly$\alpha$CoLoRe true correlation matrices achieve $\mathcal{P} \approx 0.96$ in this basis, confirming that the projection is nearly diagonal with only $\sim4\%$ of total power in the off-diagonals.

To test the sensitivity of the results to this choice of reference matrix, I repeat the full analysis using only the Ly$\alpha$CoLoRe mocks to define the reference correlation matrix. In this case the reference matrix is simply the correlation matrix from the stack of $200$ Ly$\alpha$CoLoRe mocks with $\sim1000$ sub-samples each.

\begin{figure}
    \centering
    \includegraphics[
        width=\columnwidth
    ]{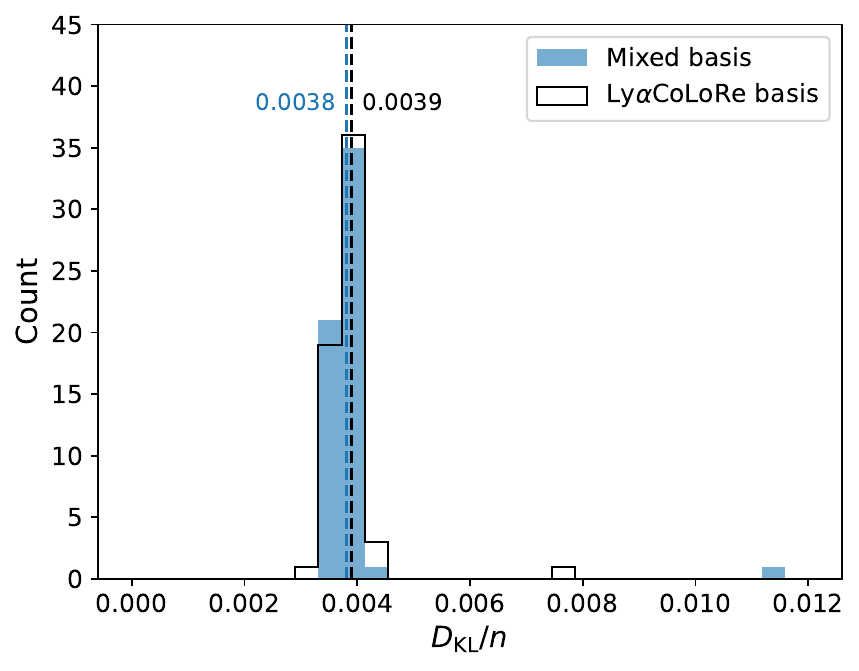}
    \caption{Distributions of the normalized KL divergence $D_\mathrm{KL}/n$ (Eq.~\ref{eq:kl}) for the final denoised covariances on the mock testing set, using the default mixed initial eigenbasis (\textit{blue filled}) and the Ly$\alpha$CoLoRe-only eigenbasis (\textit{black unfilled}). The two distributions are in close agreement, demonstrating that the results are robust to the choice of initial reference matrix.}
    \label{fig:KL_divergence_basis}
\end{figure}

Figure~\ref{fig:KL_divergence_basis} compares the $D_\mathrm{KL}/n$ distributions (Eq.~\ref{eq:kl}) after the final correction stage for the default mixed eigenbasis and the Ly$\alpha$CoLoRe-only eigenbasis. The two distributions are consistent, with the median $D_\mathrm{KL}/n$ increasing by only $\sim0.0001$ when using the Ly$\alpha$CoLoRe-only basis. This consistency is expected: noisy Ly$\alpha$CoLoRe covariances are well reconstructed by the initial stage alone, since their structure is fully captured by that eigenbasis, while noisy Saclay covariances rely more heavily on the residual correction to account for structure not represented in the initial Ly$\alpha$CoLoRe-only basis. The fact that the results remain consistent in either case demonstrates the complementary roles of the two pipeline stages.

\begin{figure*}
    \centering
    \includegraphics[
        width=0.8\textwidth
    ]{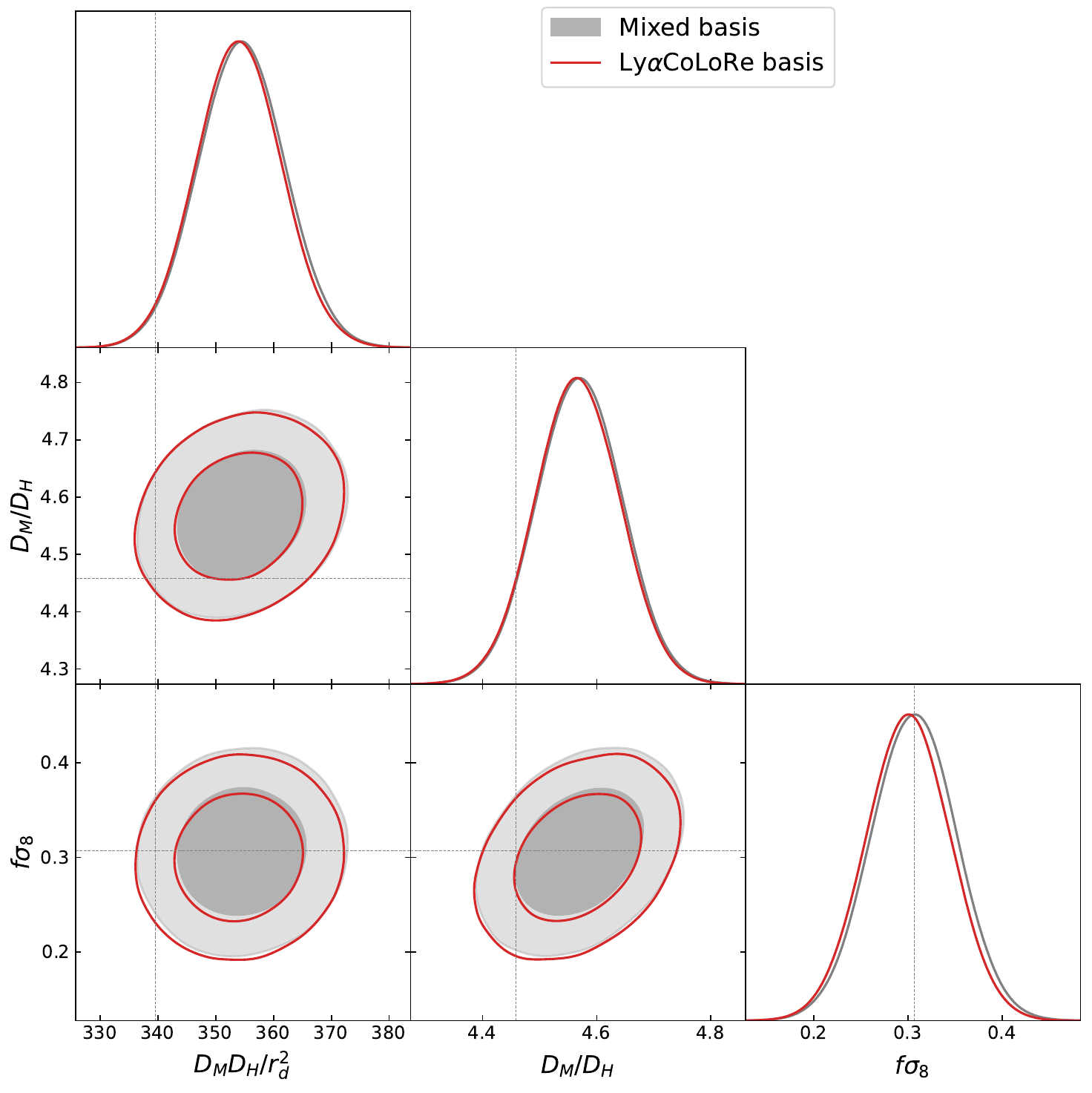}
    \caption{Gaussian posteriors of cosmological parameters from a full-shape fit on a Saclay mock in the classifier testing set. The results compare the default mixed initial eigenbasis (\textit{gray contours}; also shown in Fig.~\ref{fig:mock_chains}) and the Ly$\alpha$CoLoRe-only initial eigenbasis (\textit{red contours}). The two results are in excellent agreement, demonstrating the robustness of the denoising method to different choices of reference matrix.}
    \label{fig:mock_chains_basis}
\end{figure*}

Finally, Figure~\ref{fig:mock_chains_basis} compares the cosmological parameter constraints obtained with the two choices of reference matrix on a Saclay mock from the classifier testing set, repeating the inference shown in Figure~\ref{fig:mock_chains} but by finding the maximum likelihood solution instead of sampling the full posteriors. The results are in excellent agreement, confirming that the choice of initial reference matrix does not meaningfully affect the performance of the denoising methodology.

\end{appendix}

\end{document}